# New relations for the derivative of the confluent Heun function

**V.A. Shahnazaryan**[1,2], **T.A. Ishkhanyan**[3], **T.A. Shahverdyan**[1,3], **and A.M. Ishkhanyan**[1]

[1]Institute for Physical Research, NAS of Armenia, 0203 Ashtarak, Armenia
[2]Russian-Armenian (Slavonic) University, 0051 Yerevan, Armenia
[3]Moscow Institute of Physics and Technology, 141700 Dolgoprudni, Russia

The cases when the equation for the derivative of the confluent Heun function has only three singularities (in general, the equation has four such points) are examined. It is shown that this occurs only in three specific cases. Further, it is shown that in all these three cases this equation is reduced to some confluent Heun equation with changed parameters. This means that in these cases the derivative of the confluent Heun function is expressed via some other confluent Heun function. The analysis of the obtained relations shows that they can be useful for the construction of some exact solutions of the confluent Heun equation for specific values of involved parameters. Several examples of such cases are presented; and for all the cases the final solutions in terms of simpler special functions are explicitly presented. In addition, it is shown that in the mentioned three cases, the solution can be expanded as a series in terms of generalized hypergeometric functions of Goursat or Clausen.

The confluent Heun equation is a Fuchsian linear second order differential equation with two regular and one irregular singular points [1]. Over the last years, this equation has been attracting a considerable interest due to a number of problems encountered in mathematical physics [1] and in various areas of physics - in the theory of black holes in astrophysics [2], [3], in elementary particle physics and atomic physics [4], [5], in the quantum theory of motion of a charged particle in the field of two Coulomb centers in molecular physics [6], [7], etc. In particular, using this equation the solution of Teukolsky equation has been studied [2] and the Schrödinger equation for the hydrogen atom with a model potential of the form $V(r) = Ze^2/(r+\beta)$, where $\beta > 0$ is a parameter describing the "non-pointness" of the nucleus, has been considered [4], [5].

The construction of the theory of the confluent Heun equation presents a complicated mathematical problem, since the Frobenius power-series solution in the vicinity of a singular point does not converge at another singular point and general formulas connecting the expansions at different singularities are not known at present. Besides, there are only a few known cases where the confluent Heun function can be expressed in terms of simpler special functions of mathematical physics. Most often, such solutions are obtained from series solutions (power series [1] or series in terms of special functions [7]-[9]) by terminating the series, or by factorization of the confluent Heun equation [10]. Because of the limited range of applicability of these solutions, the construction of new solutions of different structure is of considerable interest. Another peculiarity of the confluent Heun functions is that, unlike the preceding generation of mathematical functions (hypergeometric functions, Bessel functions, etc.), its *derivatives* in the general case are not expressed in terms of other confluent Heun functions. In the present paper, however, following the approach proposed in [11], we show that in three specific cases this takes place, that is, ***there exist three cases when the derivative of the confluent Heun function is expressed in terms of another confluent Heun function with altered parameters.***

In the canonical form the confluent Heun equation is written as follows [1]:

$$u'' + \left(4p + \frac{\gamma}{z} + \frac{\delta}{z-1}\right)u' + \frac{4p\alpha z - \sigma}{z(z-1)}u = 0, \quad (1)$$

where $p, \gamma, \delta, \alpha, \sigma$ are arbitrary complex parameters (it is assumed that $p \neq 0$).

By taking the derivative of equation (1) and subtracting from the result the equation (1) multiplied by $g'/g$, where

$$g = \frac{4p\alpha z - \sigma}{z(z-1)}, \quad (2)$$

we obtain the following third-order equation:

$$u''' + \left(f - \frac{g'}{g}\right)u'' + \left(g + f' - f\frac{g'}{g}\right)u' = 0, \quad (3)$$

where $f$ is the coefficient of $u'$ in equation (1):

$$f = 4p + \frac{\gamma}{z} + \frac{\delta}{z-1}. \quad (4)$$

In the expanded form equation (3) reads

$$u''' + \left(4p + \frac{\gamma+1}{z} + \frac{\delta+1}{z-1} - \frac{4p\alpha}{4p\alpha z - \sigma}\right)u'' +$$

$$\left(-\frac{\gamma}{z^2} - \frac{\delta}{(z-1)^2} + \frac{4p\alpha z - \sigma}{z(z-1)} - \left(\frac{4p\alpha}{4p\alpha z - \sigma} - \frac{1}{z} - \frac{1}{z-1}\right)\left(4p + \frac{\gamma}{z} + \frac{\delta}{z-1}\right)\right)u' = 0. \quad (5)$$

The derived equation is a second-order equation for the confluent Heun function's derivative $u'$. Note that the presence of the term $4p\alpha/(4p\alpha z - \sigma)$ in the coefficients of this equation indicates that, as compared with the original confluent Heun equation (1), this equation generally contains an additional singularity located at the point $z = \sigma/(4p\alpha)$. However, it is evident that if this point coincides with one of the already existing singular points the number of the singularities is not increased - only the former singularities $z = 0, 1, \infty$ remain. It is not difficult to verify that the points $z = 0$ and $z = 1$ in this case still remain regular and the singularity rank of the irregular point $z = \infty$ is not changed. Evidently, this situation takes place in three cases, namely, when $\alpha = 0$, $\sigma = 0$ and $\sigma = 4p\alpha$. The immediate conclusion from this observation is that equation (5) in the indicated three cases presents some transformed confluent Heun equation for $u'$ with altered parameters as compared with the initial equation (1). Thus, for $\alpha = 0$, $\sigma = 0$, $\sigma = 4p\alpha$ the **derivative** of the confluent Heun function is expressed via some other confluent Heun function with changed parameters. In several cases, this property of the confluent Heun function can be useful. In particular, it has a potential to generate, for specific values of the involved parameters, closed solutions of the confluent Heun equation, which are expressed in terms of simpler mathematical functions. Besides, it also allows one to generate solutions in the form of a series of beta functions [12] and generalized hypergeometric functions $_2F_2$ of Goursat or $_3F_2$ of Clausen [13]. Let us consider some specific examples treating the indicated three cases separately.

1. $\alpha = 0$. In this case equation (5) is reduced to the equation

$$w'' + \left(4p + \frac{\gamma+1}{z} + \frac{\delta+1}{z-1}\right)w' + \frac{4p(2z-1)+\gamma+\delta-\sigma}{z(z-1)}w = 0, \qquad (6)$$

where we have introduced the notation $w = u'$. Hence,

$$\frac{dHC(p,\gamma,\delta,0,\sigma;z)}{dz} = HC(p,\gamma+1,\delta+1,2,\sigma+4p-\gamma-\delta;z), \qquad (7)$$

where $HC$ denotes the confluent Heun function. Now, it is clear that if $\delta = -1$ and $\sigma = 4p + \delta + \gamma$, the singularity located at $z = 1$ vanishes, and equation (6) is reduced to the confluent hypergeometric equation (Kummer's equation )

$$w'' + \left(4p + \frac{\gamma+1}{z}\right)w' + \frac{8p}{z}w = 0. \qquad (8)$$

The general solution of this equation is well known and in the general case can be represented as a linear combination of the confluent hypergeometric functions by Kummer and Tricomi [12]. For the above specific parameters the solution is further simplified and written as

$$w = C_1(4pz+\gamma-1)e^{-4pz}(-4pz)^{-\gamma} + C_2 e^{-4pz}L(1-\gamma,\gamma;4pz), \qquad (9)$$

where $L(1-\gamma,\gamma;4pz)$ is the Laguerre polynomial [12]. After integration, we obtain the following explicit solution of the initial equation (1):

$$u = \int w(z)dz = C_1 e^{-4pz}(-4pz)^{1-\gamma} + C_2\left(1 - \frac{\sigma}{4p}e^{-4pz}(-4pz)^{1-\gamma}\Gamma(\gamma,-4pz)\right), \qquad (10)$$

where $\Gamma$ is the Euler's gamma function [12]. Note that, in order to construct the correct final solution, a specific integration constant was chosen here.

2. $\sigma = 0$. In this case we have:

$$u''' + \left(4p + \frac{\gamma}{z} + \frac{\delta+1}{z-1}\right)u'' + \frac{4p(1+\alpha)z^2 + \gamma}{z^2(z-1)}u' = 0. \tag{11}$$

Applying the transformation $u' = \varphi(z)w(z)$, where $\varphi = z^s$, we obtain:

$$w'' + \left(4p + \frac{\gamma+2s}{z} + \frac{\delta+1}{z-1}\right)w' +$$

$$\frac{4p(1+s+\alpha)z^2 + s(\gamma+\delta+s-4p)z - s^2 + s(1-\gamma) + \gamma}{z^2(z-1)}w = 0. \tag{12}$$

The numerator of the coefficient of $w$ in this equation is a second-degree polynomial in $z$. In order to turn this equation into a confluent Heun equation, one should equate to zero the constant term of this polynomial, that is require $-s^2 + s(1-\gamma) + \gamma = 0$, which implies that should be $s = 1$ or $s = -\gamma$. Then, after cancelling $z$, the denominator takes the form $z(z-1)$, the numerator becomes a linear function of $z$ and we obtain a confluent Heun equation with modified parameters:

$$w'' + \left(4p + \frac{\gamma+2s}{z} + \frac{\delta+1}{z-1}\right)w' + \frac{4p(1+s+\alpha)z + s(\gamma+\delta+s-4p)}{z(z-1)}w = 0. \tag{13}$$

Thus,

$$\frac{dHC(p,\gamma,\delta,\alpha,0;z)}{dz} = z^s HC(p,\gamma+2s,\delta+1,s+\alpha+1,s(4p-\gamma-\delta-s);z), \tag{14}$$

where $s = 1$ or $s = -\gamma$.

Now, if we eliminate the singularity at the point $z = 1$ in equation (13) by an appropriate choice of the parameters, we obtain a confluent hypergeometric equation. Consider the case $s = -\gamma$. By the choice $\delta = -1$ and $\gamma = -4p(1+\alpha)$ equation (13) is reduced to

$$w'' + \left(4p + \frac{4p(1+\alpha)}{z}\right)w' + \frac{4p(1+4p)(1+\alpha)}{z}w = 0. \tag{15}$$

The solution of this confluent hypergeometric equation is

$$w = C_1 e^{-4pz} z^{1-4p(1+\alpha)} {}_1F_1(-\alpha - 4p(1+\alpha), 2 - 4p(1+\alpha); 4pz) +$$

$$C_2 e^{-4pz} z^{1-4p(1+\alpha)} L(\alpha + 4p(1+\alpha), 1 - 4p(1+\alpha); 4pz), \tag{16}$$

where $_1F_1$ is Kummer's confluent hypergeometric function [12]. Integrating, we obtain the solution of equation (1) in the explicit form:

$$u = \int z^s w(z)dz = C_1\left(z^2 G(-1,-1-\alpha,0,-1+4p(1+\alpha),-2;4pz) - \frac{\Gamma(4p(1+\alpha))}{4p\alpha\Gamma(-1-\alpha)}\right) +$$

$$C_2\left(_1F_1(\alpha,-4p(1+\alpha);-4pz) - \frac{\alpha z}{1+\alpha}{}_1F_1(1+\alpha,1-4p(1+\alpha);-4pz)\right), \quad (17)$$

where $G$ is Meijer's function [12].

Obviously, for the case $s = 1$ we will obtain a similar result.

3. $\sigma = 4p\alpha$. In this case equation (5) has the form

$$u''' + \left(4p + \frac{1+\gamma}{z} + \frac{\delta}{z-1}\right)u'' + \frac{4p(1+\alpha)(z-1)^2 - \delta}{z(z-1)^2}u' = 0. \quad (18)$$

Proceeding as in the previous case, but this time taking $\varphi = (z-1)^s$, we get

$$w'' + \left(4p + \frac{\gamma+1}{z} + \frac{\delta+2s}{z-1}\right)w' +$$

$$\frac{4p(1+s+\alpha)z^2 + (s^2 + s\gamma + s\delta - 4ps - 8p - 8p\alpha)z + 4p + 4p\alpha - s - s\gamma - \delta}{z(z-1)^2}w = 0. \quad (19)$$

The numerator of the coefficient of $w$ is a second-degree polynomial in $z$. In order to reduce this equation to the confluent Heun equation one should rewrite the numerator as a polynomial in $z-1$ and equate to zero the constant term of the resultant polynomial. The last condition is equivalent to the equation $s^2 + s(\delta - 1) - \delta = 0$, hence, $s = \{1, -\delta\}$. Equation (19) will then be transformed into the following confluent Heun equation

$$w'' + \left(4p + \frac{\gamma+1}{z} + \frac{\delta+2s}{z-1}\right)w' + \frac{4p(1+s+\alpha)z + s(s+\gamma+\delta) - 4p(1+\alpha)}{z(z-1)}w = 0. \quad (20)$$

Hence,

$$\frac{dHC(p,\gamma,\delta,\alpha,4p\alpha;z)}{dz} = (z-1)^s HC(p,\gamma+1,\delta+2s,1+s+\alpha,\sigma_1;z), \quad (21)$$

where $\sigma_1 = 4p(1+\alpha) - s(s+\gamma+\delta)$ and $s = 1$ or $-\delta$.

Now, eliminating the singularity at $z = 0$ in equation (20) by choosing specific parameters, we will arrive at a confluent hypergeometric equation. Consider the case $s = -\delta$ and assume $\gamma = -1$, $\delta = 4p(1+\alpha)$. We have

$$w'' + \left(4p - \frac{4p(1+\alpha)}{z-1}\right)w' + \frac{4p(1-4p)(1+\alpha)}{z-1}w = 0, \quad (22)$$

the solution of which is

$$w = C_1 e^{-4p(z-1)} (z-1)^{1+4p(1+\alpha)} {}_1F_1(-\alpha + 4p(1+\alpha), 2 + 4p(1+\alpha); 4p(z-1)) +$$
$$C_2 e^{-4p(z-1)} (z-1)^{1+4p(1+\alpha)} L(\alpha - 4p(1+\alpha), 1 + 4p(1+\alpha); 4p(z-1)). \qquad (23)$$

Accordingly, the solution of the confluent Heun equation (1) is:

$$u = C_1 \left( (z-1)^2 G(-1,-1-\alpha,0,-1-4p(1+\alpha),-2; 4p(z-1)) + \frac{\Gamma(-4p(1+\alpha))}{4p\alpha\Gamma(-1-\alpha)} \right) +$$
$$C_2 \left( {}_1F_1(\alpha, 4p(1+\alpha); -4p(z-1)) + \frac{\alpha(z-1)}{1+\alpha} {}_1F_1(1+\alpha, 1+4p(1+\alpha); -4p(z-1)) \right). \qquad (24)$$

Note that the comparison of equations (11) and (18) shows that the former is transformed into the latter by the replacement $z \to 1-z$, $\gamma \leftrightarrow \delta$ and $p \to -p$. Therefore, the solution (24) can be obtained from equation (17) by the mentioned transformation. The verification shows that this is indeed the case, so that the consideration can always be confined to the case $\sigma = 0$ only.

Let us now show that the derived relations for the derivative of the confluent Heun function can also generate solutions of a different structure. The simplest possibility arises if we construct the solution of equation (13) for $w$ in the form of the Frobenius series. Indeed, it is easy to see that in that case after multiplication by $\varphi = (z-1)^s$ and integration, the solution of the initial confluent Heun equation (1) is written in the form of a series in terms of incomplete beta functions: $B_z(a,b) = \int_0^z t^{a-1}(1-t)^{b-1} dt$, $\text{Re}(a) > 0$. Note, that such an expansion of the solution of the confluent Heun equation in terms of beta functions is an alternative way to reproduce the results of [9]. Further, note that the solution of the equation (13) for the derivative in certain cases can be constructed as a series in terms of beta functions [9]. Then, as a result of integration, the solution of equation (1) will be written as a series involving certain two-term combinations of the products of elementary and beta functions. Other types of solutions can be constructed if we start from the solution of equation (13) as a series in terms of hypergeometric functions. For example, if a series in terms of Kummer's functions ${}_1F_1$ is taken, then the solution of the initial equation (1) will involve Goursat's generalized hypergeometric functions ${}_2F_2$ [13]. In the case when a solution of equation (13) as a series in terms of the Gauss hypergeometric functions ${}_2F_1$ is used, the solution of the equation (1) will involve Clausen's generalized hypergeometric functions $_3F_2$ [13].

Consider an example of the solution in the form of a series in terms of the Goursat functions. For $s = -\gamma$ equation (13) is rewritten as

$$w'' + \left(4p - \frac{\gamma}{z} + \frac{\delta+1}{z-1}\right)w' + \frac{4p(1+\alpha-\gamma)z - \gamma\delta + 4p\gamma}{z(z-1)}w = 0. \qquad (25)$$

We seek for the solution of this equation in the form

$$w = \sum_n a_n w_n = \sum_n a_{n1} F_1(\alpha_n; \gamma_n; s_0 z), \qquad (26)$$

where $_1F_1(\alpha_n; \gamma_n; s_0 z)$ is the solution of the following confluent hypergeometric equation:

$$w_n'' + \left(\frac{\gamma_n}{z} - s_0\right)w_n' - \frac{\alpha_n s_0}{z} w_n = 0. \qquad (27)$$

Substituting (26) in (25) and using (27), we have:

$$\sum_n a_n \left[\left(4p + s_0 - \frac{\gamma + \gamma_n}{z} + \frac{\delta+1}{z-1}\right)w_n' + \frac{4p(1+\alpha-\gamma)z - \gamma\delta + 4p\gamma + \alpha_n s_0(z-1)}{z(z-1)}w_n\right] = 0, \qquad (28)$$

or

$$\sum_n a_n \left[\left((4p + s_0)z(z-1) - (\gamma + \gamma_n)(z-1) + (\delta+1)z\right)w_n' + \right.$$
$$\left. \left((4p(1+\alpha-\gamma) + \alpha_n s_0)z - \gamma\delta - \alpha_n s_0 + 4p\gamma\right)w_n\right] = 0 \qquad (29)$$

To proceed further, it is now necessary to apply the recurrence relations between the involved confluent hypergeometric functions. We choose the parameters as $\alpha_n = \alpha_0 + n$, $\gamma_n = \gamma_0 + n$ and note that

$$w_n' = s_0 \frac{\alpha_n}{\gamma_n} w_{n+1}, \quad z(w_n' - s_0 w_n) = (\gamma_n - 1)(w_{n-1} - w_n). \qquad (30)$$

To eliminate the dependence on $z^2$ in equation (29) we set $s_0 = -4p$. The equation will then get the form:

$$\sum_n a_n \left[((1+\delta-\gamma-\gamma_n)z + \gamma + \gamma_n)w_n' + (4p(1+\alpha-\gamma-\alpha_n)z - \gamma\delta + 4p(\gamma+\alpha_n))w_n\right] = 0. (31)$$

It is easy to note that for $\alpha_0 = \alpha + \gamma_0 - \delta$ it holds

$$4p(1+\alpha-\gamma-\alpha_n) = A(1+\delta-\gamma-\gamma_n), \qquad (32)$$

whence: $A = 4p = -s_0$. Thus, we get:

$$\sum_n a_n \left[(1+\delta-\gamma-\gamma_n)z(w_n' + 4pw_n) + (\gamma+\gamma_n)w_n' + (-\gamma\delta + 4p(\gamma+\alpha-\delta+\gamma_n))w_n\right] = 0. (33)$$

Using the recurrence relations (30), we obtain:

$$\sum_n a_n \left[ (1+\delta-\gamma-\gamma_n)(\gamma_n-1)(w_{n-1}-w_n) - 4p(\gamma+\gamma_n)\frac{\alpha_n}{\gamma_n} w_{n+1} + \right.$$
$$\left. (-\gamma\delta + 4p(\gamma+\alpha-\delta+\gamma_n))w_n \right] = 0. \quad (34)$$

Thus, the recurrence relation for the coefficients of expansion (26) has the form:

$$R_n a_n + Q_{n-1} a_{n-1} + P_{n-2} a_{n-2} = 0, \quad (35)$$

where

$$R_n = (1+\delta-\gamma-\gamma_n)(\gamma_n-1), \quad (36)$$

$$Q_n = 4p(\gamma+\alpha-\delta+\gamma_n) - \gamma\delta - (1+\delta-\gamma-\gamma_n)(\gamma_n-1), \quad (37)$$

$$P_n = -4p(\gamma+\gamma_n)\frac{\alpha_n}{\gamma_n}. \quad (38)$$

For left-side termination of the derived series (say, when $n=0$), should be $a_{-2}=a_{-1}=0$ and $R_0=0$. The latter takes place only when

$$\gamma_0 = 1+\delta-\gamma, \quad \alpha_0 = 1+\alpha-\gamma. \quad (39)$$

Hence, the final solution of equation (13) is explicitly written as

$$w = \sum_{n=0}^{\infty} a_n {}_1F_1(1-\gamma+\alpha+n; 1-\gamma+\delta+n; -4pz). \quad (40)$$

Accordingly, a solution of the initial equation (1) is (up to a constant factor):

$$u = \int z^s w(z) dz = C_0 + z^{1-\gamma} \sum_{n=0}^{\infty} a_n {}_2F_2(1-\gamma, 1-\gamma+\alpha+n; 2-\gamma, 1-\gamma+\delta+n; -4pz), \quad (41)$$

where, in order to satisfy the Heun equation (1), a specific value of the integration constant $C_0$ should be chosen. Finally, note that the second independent expansion can be constructed in the same manner using the previously transformed equation (25) by the change of the independent variable $z \to 1-z$.

The derived series terminates when two consecutive coefficients become equal to zero: $a_N = a_{N+1} = 0$ ($a_{N-1} \neq 0$). This is equivalent to two conditions, one of which is reduced to the equation $P_{N-1} = 0$. From equations (38) and (39) we find that this equation is fulfilled if $\delta = -N$ or $\alpha-\gamma = -N$. Note that in the latter case the Goursat functions become polynomials. As regards the second necessary condition, it leads to a $N$-th order polynomial equation for $p$, which means that generally there are $N$ cases of termination of the series (41). Taking into account the two sets emerging from the first condition, we see that the total number of possible cases of series termination for a given $N$ is $2N$. A final remark is that the direct verification reveals that at termination of the series the constant $C_0$ of equation (41) vanishes.

Thus, we have studied the properties of the linear differential equation satisfied by the derivative of the confluent Heun function. We have shown that in three specific cases, namely, when $\alpha = 0$, $\sigma = 0$, $\sigma = 4p\alpha$, this equation has only 3 singular points and by transformation of the dependent variable can be reduced to some confluent Heun equation with changed parameters. Consequently, in these three cases the derivative of the confluent Heun function is expressed via some other confluent Heun function with altered parameters [see relations (7), (14) and (21)]. It should be noted that the properties of the confluent Heun function's derivative have been discussed by several authors (see, e.g., [14]), however, the approach applied here as well as the derived relations differ from those reported previously.

We have shown that this property of the confluent Heun functions in certain cases can generate, for the confluent Heun equation, closed solutions expressed via more familiar special functions. For all three cases we have presented specific examples by constructing the explicit final solutions of the initial equation and by indicating the appropriate values of the involved parameters. We note that the constructed solutions in general differ from the previously found exact solutions obtained using other methods, in particular, by terminating the power series [1] and series in terms of special functions [7]-[9] or by factorization of the initial confluent Heun equation [10]. In addition, using as a solution of equation (13) a series in terms of confluent hypergeometric functions $_1F_1$, we have constructed a solution of the initial confluent Heun equation as a series in terms of generalized hypergeometric Goursat functions $_2F_2$. We have also shown that in the mentioned three cases the solution can be expanded as a series in terms of Clausen's generalized hypergeometric functions $_3F_2$.

An interesting feature of the constructed closed-form exact solutions is that in all three cases at least one of the fundamental solutions presents a two-term combination [see the equations (10), (17) and (24)]. Solutions of such a structure were previously mentioned in literature, for instance, when studying the exact solutions of the one-dimensional stationary Schrödinger equation in terms of the Airy functions for the quark-antiquark interaction with the potential $U = V_0/x^{3/2} - 5/(36x^2)$ [15]. As the solution of the corresponding problem shows, in these cases the dependences of the bound states' energy spectrum on the principal quantum number sometimes reveal features that have no known analogues. This observation motivates us to examine the physical problems described by the confluent Heun functions discussed above. We hope to present the corresponding results in the near future.

**Acknowledgments**

This research has been conducted within the scope of the International Associated Laboratory (CNRS-France & SCS-Armenia) IRMAS. The research has received funding from the European Union Seventh Framework Programme (FP7/2007-2013) under grant agreement No. 205025 – IPERA. The work has been supported by the Armenian National Science and Education Fund (ANSEF Grant No. 2591) and Armenian State Committee of Science (SCS Grant No. 11RB-026). The authors express special thanks to Prof. Artur Ishkhanyan for formulation of the research topic and permanent assistance during the research.